# Clustering of the Ly$_\alpha$ Clouds in the Line of Sight to the z=3.66 QSO 0055-269

S. Cristiani[1], S. D'Odorico[2], A. Fontana[3], E. Giallongo[3] and S. Savaglio[4]

[1] Dipartimento di Astronomia, Università di Padova, vicolo dell' Osservatorio 5, I-35122 Padova, Italy

[2] European Southern Observatory, Karl Schwarzschild Str. 2, D-85748 Garching, Germany

[3] Osservatorio Astronomico di Roma, via dell'Osservatorio, I-00040 Monteporzio, Italy

[4] Dipartimento di Fisica, Università della Calabria, I-87036 Arcavata di Rende, Cosenza, Italy

**Abstract.** The spectrum of the Q0055-269 ($z = 3.66$) has been observed at the resolution of 14 km s$^{-1}$ in the wavelength interval 4750–6300 Å. The statistical distribution of the Doppler parameter for the Ly$_\alpha$ lines is peaked at $b \simeq 23$ km s$^{-1}$. The column density distribution is described by a power-law with a break at $\log N_{HI} \simeq 14.5$. Significant clustering, with $\xi \simeq 1$ at $\Delta v = 100$ km s$^{-1}$, is detected for lines with $\log N_{HI} \geq 13.8$. Two voids of size $\sim 2000$ km s$^{-1}$ are found in the spectrum with a probability of $2 \times 10^{-4}$.

## 1 The clustering properties of the Ly$_\alpha$ clouds

No clustering in the velocity space has been detected so far for Ly$_\alpha$ lines on scales $300 < \Delta v < 30000$ km s$^{-1}$ (Sargent et al. 1980, 1982; Webb & Barcons 1991). Their spatial distribution is different from that of the metal-line systems selected by means of the CIV doublet, that are known to cluster on scales at least up to 600 km s$^{-1}$ (Sargent et al. 1988). Preliminary results at higher resolution seem to indicate weak clustering on smaller scales ($\Delta v = 50 - 300$ km s$^{-1}$, Webb 1987, Chernomordik, this conference).

The larger the density of the observed lines, the better the chances of detecting a clustering signal. For this reason we have looked for a QSO of redshift 3.66, Q0055-269, which has been observed at ESO La Silla with the NTT telescope and the EMMI instrument in the echelle mode (Cristiani et al., 1995). The weighted mean of the flux-calibrated spectra has a resolution $R \sim 22000$. The signal-to-noise ratio per resolution element at the continuum level ranges from $S/N = 12$ to 40 in the interval 4750–6300 Å.

Column densities and Doppler widths of the Ly$_\alpha$ lines and metal-line systems have been derived by a fitting procedure. The statistical distribution of the Doppler parameter for the Ly$_\alpha$ lines is peaked at $b \simeq 23$ km s$^{-1}$. The column density distribution is described by a power-law with a break or cutoff at $\log N_{HI} \simeq 14.5$. A featureless distribution is rejected with a probability of 99.94%.

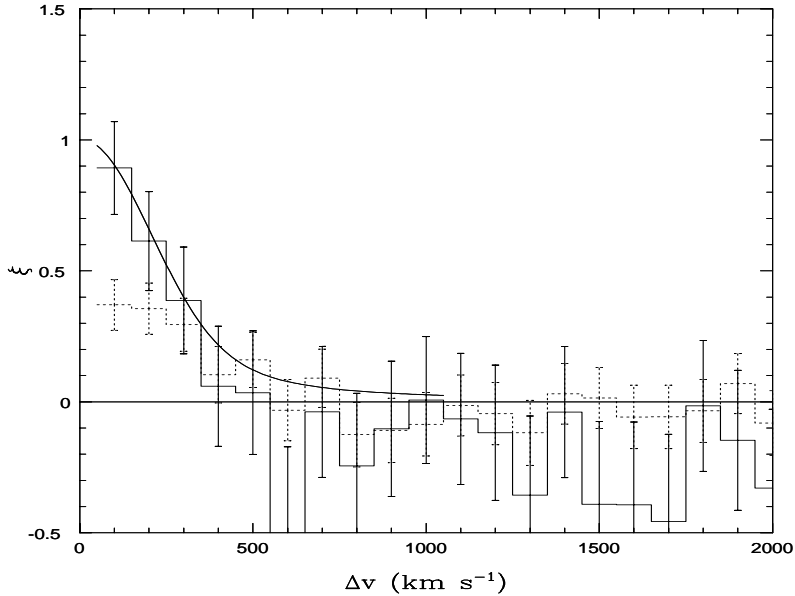

**Fig. 1.** Two-point correlation function for the Lyman−α lines in Q0055-269. The continuous histogram is for lines with $\log N_{HI} \geq 13.8$; the dotted histogram is for lines with $\log N_{HI} \geq 13.3$. The continuous curve is the model described in Sec. 2 and computed from the eq. (2) with $\gamma = 1.77$, $\sigma = 150$ km s$^{-1}$, $r_{cl} = 110$ kpc and $r_o = 280$ kpc at $z = 3.3$.

The two-point correlation function in the velocity space is defined as

$$\xi(v, \Delta v) = \frac{N_{obs}(v, \Delta v)}{N_{exp}(v, \Delta v)} - 1 \qquad (1)$$

where $N_{obs}$ is the observed number of line pairs with velocity separations between $v$ and $v+\Delta v$ and $N_{exp}$ is the number of pairs expected from a random distribution in redshift. In our line sample $N_{exp}$ is obtained averaging 1000 numerical simulations generated according to the cosmological distribution $\propto (1+z)^\gamma$, derived from the maximum likelihood analysis of the real line sample in the given interval of column density and redshift. In this way it is possible to correct for incomplete wavelength coverage due to gaps in the spectrum or occultation of weak lines due to strong complexes. No velocity splittings $\Delta v < 20$ km s$^{-1}$ were included because of the intrinsic line blending due to the typical widths of the Ly$_\alpha$ lines. The $1\sigma$ deviation from a random distribution is given by the $N_{obs}^{-1/2}$, a good approximation in the case of weak clustering $\xi \leq 1$. The resulting correlation function of Ly$_\alpha$ lines in Q0055-269 with $N_{HI} > 13.3$ is shown in Figure 1. It is clear that a weak but significant signal is present with $\xi \simeq 0.34 \pm 0.06$ up to 350 km s$^{-1}$.

Exploring the variations of the correlation function as a function of the column density, an increase in the first velocity bin appears as the col-umn density threshold is raised. The maximum significant signal is obtained for lines with $\log N_{HI} \geq 13.8$, for which the correlation function is $\xi = 0.89 \pm 0.18$ for $\Delta v = 100$ km s$^{-1}$. When computing $\xi$ for lines in the interval $\log N_{HI} = 13.3 - 13.6$ (including in the simulated spectra gaps due to lines with $\log N_{HI} > 13.8$), no significant clustering is observed.

For comparison the correlation function of the Ly$_\alpha$ samples obtained from the QSO PKS 2126-158 (Giallongo et al., 1993) and 0014+81 (Rauch et al. 1992) has been computed. A significant clustering for $\Delta v = 100$ km s$^{-1}$ appears when the samples are limited to strong lines (with a lower significance due to the smaller density of lines in the individual spectra): in PKS 2126-158 $\xi = 1.02 \pm 0.26$ is obtained for lines with $\log N_{HI} \geq 13.8$, in Q0014+81 $\xi = 0.85 \pm 0.30$ for lines with $\log N_{HI} \geq 14.1$.

The correlation found for the Ly$_\alpha$ cloud positions is less pronounced than for metal-line absorption systems or galaxies, but consistent with a scenario of gravitationally induced correlations, as expected in models where gravitation is an important confining agent (Mo, Miralda-Escudé, & Rees 1993). The correlation found would imply, in the standard CDM framework, that the Ly$_\alpha$ clouds responsible for the weakest absorption are abundant in underdense regions. The redshift autocorrelation function could be lower than the spatial autocorrelation.

## 2 The dimensions of the Ly$_\alpha$ clouds

At small velocity differences the three-dimensional spatial autocorrelation function $\xi_r$ and the redshift autocorrelation function are related by the convolution (Heisler, Hogan, & White, 1989):

$$\xi_v = \int_0^\infty H dr\, \xi(r)\, P(v \mid r) \propto \int_{r_{cl}}^\infty \frac{H dr}{\sigma} \left(\frac{r}{r_o}\right)^{-\gamma} \left[e^{-\frac{(Hr-v)^2}{2\sigma^2}} + e^{-\frac{(Hr+v)^2}{2\sigma^2}}\right] \quad (2)$$

where a Gaussian distribution of the peculiar motions with respect to the Hubble flow is assumed together with a power-law spatial correlation function of the galaxy-type.

At small velocity splittings, the redshift correlation scale $r_o$ depends mainly on the cloud sizes $r_{cl}$ and on the velocity dispersion. Although these quantities are poorly known, it is instructive to derive constraints on the cloud sizes and velocity dispersions from the observed correlation. Assuming an index $\gamma = 1.77$ for the power-law spatial correlation function, the fit shown in Fig. 1 provides upper limits on the cloud sizes as a function of the velocity dispersion and of the appropriate correlation scale. We obtain $r_{cl} \sim 100 - 190$ kpc for $\sigma_v = 50 - 160$ km s$^{-1}$, and $r_o \sim 245 - 420$ kpc, respectively. In the former case a good fit to the overall shape of the observed correlation function requires a very low value $\gamma = 1.1$. For $\sigma_v > 200$ km s$^{-1}$ the redshift correlation function becomes too flat and very steep $\gamma$ values are needed (e.g. $\sigma_v = 300$ km s$^{-1}$ requires $\gamma = 4$).

## 3    Voids in the Ly$_\alpha$ forest

Voids in the Ly$_\alpha$ forest provide a test for models of the large-scale structure, even if changes in the IGM due to fluctuations of the UV ionizing flux can be efficient in depleting neutral hydrogen along the line of sight.

Searches for megaparsec-sized voids have produced a couple of claims: a void with comoving size $\sim 30$ Mpc in the spectrum of Q0420-388 (Crotts 1987,1989), whose statistical significance has been questioned (Ostriker, Bajtlik, & Duncan 1988); another one in Q0302-003, with a size $\sim 23$ Mpc (Dobrzycki & Bechtold 1991). The statistical significance of a given void depends exponentially on the line density (Ostriker et al. 1988) and uncertainties in the line statistics strongly influence the probability estimate. High resolution data, less affected by blending effects, are ideal in this respect.

We have searched for gaps in a sample of Ly$_\alpha$ lines of Q0055-269 limited, to ensure statistical completeness and accuracy in the deblending, to $\log N_{HI} \geq 13.3$ and $b \leq 30$ km s$^{-1}$. Two regions devoid of such lines have been found, centered respectively at $\lambda \sim 5000, 5206$ Å, with sizes $\Delta v \sim 2009, 2046$ km s$^{-1}$ (i.e. $\sim 20$ Mpc). To establish the random probability of observing such gaps, a set of 50000 spectra have been simulated with the observed number ($N = 178$) of redshifts randomly generated in the same interval of the data (excluding the region within 8 Mpc from the quasar) according to the cosmological distribution. The probability of getting a gap larger than the maximum observed one is then $\simeq 0.02$. The 2046 km/s gap is significant only to $2.2\sigma$ level, but the joint probability for the presence of the two observed gaps in the same spectrum is $\simeq 2 \times 10^{-4}$. Even if underdense regions are statistically significant in our spectrum, the filling factor is less than 10%.